\documentclass[epsfig,aps,pra,eqsecnum,twocolumn]{revtex4}
\usepackage{graphicx}

\newcommand{\be}{\begin{equation}}
\newcommand{\ee}{\end{equation}} 
\newcommand{\ba}{\begin{eqnarray}}
\newcommand{\ea}{\end{eqnarray}}

\def\opone{\leavevmode\hbox{\small1\kern-3.5pt\normalsize1}}
\def\l{\label}

\def\r{\ref}
\def\la{\langle}
\def\ra{\rangle}
\def\e{{\rm e}}
\def\i{{\rm i}}
\def\f{\frac}
\def\s{\sqrt}

\begin{document}

\title{Decoherence scenarios from micro- to macroscopic superpositions}

\author{Walter T. Strunz}
\affiliation{Fakult\"at f\"ur Physik, 
Universit\"at Freiburg, Hermann-Herder-Str.3,
79104 Freiburg, Germany}
\author{Fritz Haake}
\affiliation{Fachbereich Physik, Universit\"at Essen,
45117 Essen, Germany}


\date{\today}

\begin{abstract}



Environment induced decoherence entails the absence of quantum interference
phenomena from the macroworld.
The loss of coherence between superposed wave packets
is a dynamical process  the speed of which depends on the packet separation:
The farther the packets are apart, the faster they decohere.
The precise temporal course depends on the relative size of the time scales
of decoherence and other processes taking place in the open system and its 
environment. We use the exactly solvable model of an harmonic oscillator 
coupled to a bath of harmonic oscillators to illustrate various decoherence 
scenarios: These range from exponential
golden-rule decay for microscopic superpositions,
system-specific decay for larger separations in a crossover regime,
and finally the universal interaction-dominated decoherence for
ever more macroscopic superpositions
investigated in great generality in the accompanying paper 
[W. T. Strunz, F. Haake, and D. Braun; Phys. Rev. A, henceforth
referred to as [SHB]].

\end{abstract}

\maketitle

\section{Introduction}

A collection of $N+1$ harmonic oscillators $N$ of which are mutually
free but all coupled, symmetrically and harmonically, to the remaining 
oscillator, enjoys considerable popularity as a model of an $N$-freedom 
environment acting on a single-freedom system 
\cite{generalosci,masterosciref}.  The importance of the 
model lies in its rigorous explicit tractability through a normal-mode 
analysis. Most applications aim at revealing the effectively irreversible 
behavior of the central oscillator brought about by the
coupling to the environment (alias ``bath'') for large $N$; 
in the limit $N\to\infty$ even strict irreversibility results under 
certain assumptions for the distribution 
of bath frequencies, and the central oscillator becomes linearly damped.

While obliged to the tradition just pointed to, the present paper picks 
up a more recent trend and exploits the rigorous tractability of the 
model to reveal the emergence of classical behavior from quantum 
dynamics in the passage from the microscopic
to the macroscopic world
\cite{Zeh,Zurek,Habib}. In particular, we study the temporal fate of 
superpositions of two wave packets for the central oscillator
as in \cite{Habib}, yet concentrate on the various decoherence
scenarios emerging 
as the distance of the superposed wave packets is varied
from microscopic to macroscopic scales.
As is by now well known such superposed packets
loose their relative coherence, due to the dissipative influence of the 
bath, the faster the larger the initial separation of the two wave 
packets. The life time $\tau_{\rm dec}$ of the relative 
coherence is  inversely proportional to a power of the initial 
separation, $\tau_{\rm dec}\propto (\lambda/d)^{\nu}$ with $\nu>0$; 
within the power law, the 
separation $d$ is referred to a microscopic quantum scale $\lambda$, and 
therefore exceedingly
rapid decoherence results as $d$ is increased towards macroscopic magnitude. 
Keeping all other parameters of the problem fixed, it is the
initial separation between the two wave packets only that determines the 
size of the decoherence time scale
$\tau_{\rm dec}$ relative to other relevant system or environmental time
scales.

The most familiar golden-rule limit
\begin{equation}
\tau_{\rm sys}\ll\tau_{\rm dec}\ll \tau_{\rm diss}\,,
\label{1.1}
\end{equation}
which allows the system to undergo many cycles during decoherence,
can hold only as long as the separation between the two
wave packets remains below a certain limit. On increasing the
separation we encounter a qualitatively different
regime in which decoherence is faster than any system time scale,
\begin{equation}
  \tau_{\rm dec}\ll\tau_{\rm sys}, \tau_{\rm diss},
  \label{1.2}
\end{equation}
irrespective of the relative size of $\tau_{\rm sys}$ and $\tau_{\rm
diss}$. In that {\it interaction dominated limit} the free-motion 
Hamiltonian of the
central oscillator, rather than the interaction with the bath, behaves like a 
weak perturbation during decoherence. For yet larger separations 
decoherence becomes 
the fastest process by far, faster even than reservoir time scales,
\begin{equation}
  \tau_{\rm dec}\ll\tau_{\rm res}, \tau_{\rm sys}, \tau_{\rm diss}\,.
  \label{1.3}
\end{equation}
The universal behavior resulting in the limit
(\ref{1.3}) may look like instantaneous decoherence on the classical 
time scales $\tau_{\rm sys},\tau_{\rm diss}$. While certainly requiring 
separations $d$ huge on the quantum scale $\lambda$, the limit (\ref{1.3}) 
will turn out to allow, surprisingly, moderate 
or even small $d$ relative to every-day macroscopic scales.

The pure limiting cases mentioned above allow for analytical treatment 
for general open systems, as shown in the accompanying paper [SHB]. Crucially,
decoherence in the interaction dominated cases (\ref{1.2}) or
(\ref{1.3}) becomes independent of the system Hamiltonian and
may thus be regarded as the origin of the universally observed
absence of quantum interferences in the macroworld. All of these general 
findings
of the accompanying paper will be illustrated for the oscillator model here. 
However, the principal purpose of the present paper is to study the 
interesting crossovers between the three regimes mentioned; since these 
crossover effects elude the general asymptotic
methods of [BHS], the exact tractability of the oscillator model allows 
precious insights into the emergence of classical behavior.

\section{Decoherence of superposed wave packets}\label{norm12}

We illustrate the various decoherence scenarios for a superposition of
Gaussian wave packets
\be
|\varphi\ra=c_1|\varphi_1\ra+c_2|\varphi_2\ra\,,\quad |c_1|^2+|c_2|^2=1.
\label{2.1}
\ee

In the position representation,
\be
\la q|\varphi_i\ra=\varphi_i(q)=\f{1}{(2\pi\sigma)^{1/4}}\,\e^{\i
p_i(q-q_i)/\hbar}
\,\e^{-(q-q_i)^2/4\sigma}
\label{2.2}
\ee
with $\quad i=1,2\,$.
These packets are located in position space at $q_i$ with (rms)
uncertainty $\Delta q=\s{\sigma}$ and in momentum space at $p_i$ with
uncertainty $\Delta p=\hbar/2\s{\sigma}$.
We choose coherent states \cite{Glauber} with the minimum
uncertainty
$\Delta q\Delta p=\hbar/2$ such that
both $\Delta q$ and $\Delta p$ are
$\propto\s{\hbar}$.
To ensure good separation
we stipulate that either $\Delta q\ll|q_1-q_2|$ or $\Delta
p\ll|p_1-p_2|$ or both (see Fig. 1).

\begin{figure}[h]\label{fig1}
\includegraphics[angle=270,scale=.30]{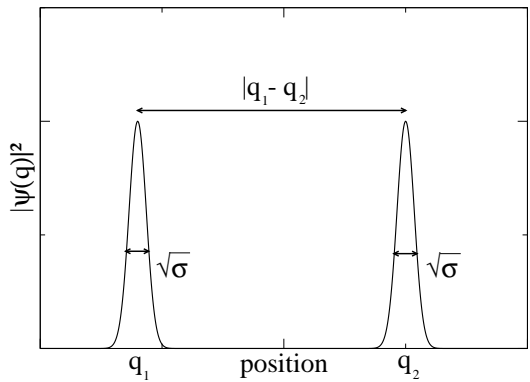}
\caption{Position space density $|\psi(q)|^2$ for a
superposition of two Gaussian wave packets with mutual distance much
larger than individual spread.}
\end{figure}

The initial density operator corresponding to the state (\ref{2.1}) is a sum 
of four terms,
\be
\rho_{\rm sys}(0)=\sum_{i,j=1}^2c_ic_j^*\,|\varphi_i\rangle\langle\varphi_j|
=\sum_{i,j}c_ic_j^*\,\rho^{ij}_{\rm sys}(0)\,,
\l{2.3}
\ee
two ``diagonal'' ones weighted by probabilities $|c_i|^2$ and two
off-diagonal ``interference terms'' $\rho^{12}_{\rm
sys}(0)=|\varphi_1\ra\la\varphi_2|=\rho^{21}_{\rm sys}(0)^\dagger$
weighted by the ``coherences'' $c_1c_2^*$ and $c_1^*c_2$.

As in [SHB] we may employ the norm
\be
N_{12}(t)={\rm Tr_{sys}}\, \rho^{12}_{\rm sys}(t)\rho^{12}_{\rm sys}(t)^\dagger
\l{2.4}
\ee
as an indicator of the temporal fate
of the relative coherence between the two superposed wave packets. Clearly, if 
the system were closed its unitary time evolution would leave that norm 
constant in time, $N_{12}(t)=1$; interaction with a many-freedom environment 
will cause decay.

\section{Harmonic-Oscillator Model}\label{Secthomodel}

Our aim is to 
illustrate various decoherence scenarios for quantum
superpositions $|\varphi\rangle = c_1|\varphi_1\rangle+ c_2|\varphi_1\rangle$
of wave packets of a harmonic oscillator 
of mass $M$ and frequency $\Omega$.
Position and momentum operators $Q$ and
$P$ obey the usual commutation relation $[Q,P]=\i\hbar$.
The reservoir is a
collection of harmonic oscillators as well, the $i$th of which has
the coordinate $Q_i$ and the momentum $P_i$, frequency $\omega_i$, and
mass $m$; the coupling is taken bilinear in the positions, such that the
three terms in the Hamiltonian $H= H_{\rm sys}+ H_{\rm res}+ H_{\rm int}$ 
read
\ba
H_{\rm sys} &=& \frac{P^2}{2M} + \frac{1}{2}M\Omega^2 Q^2\,,\nonumber\\
H_{\rm res} &=& \sum_{i=1}^N\left(\frac{P_i^2}{2m} +
\frac{1}{2}m\omega_i^2 Q_i^2\right)\,,\l{7.1}\\
H_{\rm int} &=& QB = Q\sum_{i=1}^N g_i Q_i\,. \nonumber
\ea
This model and variants thereof
have been used extensively over the years
to investigate dissipative
quantum dynamics \cite{generalosci,masterosciref,feynman}, and decoherence in
particular \cite{unruhetc}. Its popularity is due to the fact that it allows 
for an explicit exact solution for the many-body Schr\"odinger equation.

The dissipative and 
decohering influence of the reservoir is encoded in the thermal
autocorrelation function of the bath coupling agent $B$,
\begin{equation}
\langle \tilde{B}(t)\tilde{B}(0)\rangle=
\sum_i \frac{\hbar g_i^2}{2m\omega_i}
 \left[(2n_{\rm th}(\omega_i) + 1) \cos \omega_i t -\i \sin \omega_i t\right],
\l{6.2}
\end{equation}
where the time dependence in $\tilde{B}(t)=\e^{\i H_{\rm res}t/\hbar}B
\e^{-\i H_{\rm res}t/\hbar}$ refers to the free motion of the bath and
$n_{\rm th}(\omega) = (\e^{\hbar\omega/kT}-1)^{-1}$ is the
thermal number of quanta in an oscillator with angular
frequency $\omega$.
Assuming the number $N$ of bath oscillators to be large, it is customary 
\cite{Weiss} to introduce  the spectral density 
\begin{equation}
J(\omega) \equiv \frac{\pi}{2} 
 \sum_i \frac{g_i^2}{m\omega_i} \delta(\omega-\omega_i)
\end{equation}
such that the real and imaginary parts of
the correlation may be expressed as
\begin{eqnarray} \label{bathcorr} \nonumber
 \langle\frac{1}{2}\{\tilde{B}(t),B\}\rangle
& = & \frac{\hbar}{\pi}\int_0^\infty d\omega J(\omega)
 (2n_{\rm th}(\omega) + 1) \cos \omega t
\\
 \langle\frac{\i}{\hbar}[\tilde{B}(t),B]\rangle
& = & \frac{2}{\pi}\int_0^\infty d\omega J(\omega) \sin \omega t.
\end{eqnarray}
Note that the imaginary part (the damping kernel) is independent
of $\hbar$, while the real part, which describes equilibrium fluctuations, 
becomes independent of $\hbar$ only in the high temperature 
limit when $kT/\hbar$ is the largest frequency involved. 
For small temperatures, however, only quantum fluctuations remain, 
such that the real part of the correlation function becomes of first order in 
$\hbar$.
A so-called Ohmic bath is provided by a spectral density with
a linear frequency dependence for small $\omega$,
\begin{equation} \label{Ohm}
J(\omega) = M\gamma\omega f_c(\omega/\Lambda)\,
\end{equation}
with a cutoff function such that $f_c(0)=1$.
The rate $\gamma$ is a measure of the coupling strength and
turns out to be
the classical damping rate;
$\Lambda$ is a cutoff frequency. For the model to be physically
sensible, the cutoff frequency $\Lambda$ is assumed much larger than the
frequency $\Omega$ and the damping constant $\gamma$ such that
\begin{equation}
  f_c(\Omega/\Lambda)\approx 1 \,.
  \label{cutoff}
\end{equation}
We shall use $f_c(x) = 1/(1+x^2)^2$ for our simulations. 
Our model gives an initial value
$\langle B^2 \rangle =
\frac{\hbar M \gamma}{\pi}\int_0^\infty d\omega \omega
 (2n_{\rm th}(\omega) + 1) f_c(\omega/\Lambda)$. 
In the high-temperature limit, we obtain
\begin{equation}\label{bquadrat}
\langle B^2 \rangle = M kT \gamma\Lambda \frac{2}{\pi}\int_0^\infty dx f_c(x)
 = M kT \gamma\Lambda \cdot {\cal O}(1)\,,
\end{equation}
the remaining integral being a real number of order one. The zero-temperature 
limit
\begin{equation}
\langle B^2 \rangle = \hbar M \gamma\Lambda^2 \frac{1}{\pi}
\int_0^\infty dx x f_c(x) =
 \hbar M \gamma\Lambda^2 \cdot {\cal O}(1)
\end{equation}
mainly differs from the high-temperature one by the replacement of the thermal
energy $kT$ with the cutoff energy $\hbar\Lambda$. The choice of frequencies
$\Omega, \gamma, \Lambda, kT/\hbar$ determines the system and
reservoir time scales of our model.
We choose $\gamma = 10^{-5}\Omega$, $\Lambda = 10^2\Omega$,
$kT/\hbar = 20 \Omega$.

\subsection{Dynamics: Exact Master equation}

The dynamics of the system is determined by the reduced density 
operator $\rho_{\rm sys}(t)\equiv\rho_t$.
Assuming initial decorrelation of system and bath, we encounter
the following well known evolution equation \cite{masterosciref}
\begin{eqnarray}\label{masterosci}
\dot\rho_t & = & \frac{1}{\i\hbar} [H_{\rm sys},\rho_t] \\ \nonumber
& & +  \frac{a_t}{2\i\hbar} [Q^2,\rho_t]
+ \frac{b_t}{2\i\hbar} [Q,\{P,\rho_t\}] \\ \nonumber
& & + \frac{c_t}{\hbar^2} [Q,[P,\rho_t]]
- \frac{d_t}{\hbar^2} [Q,[Q,\rho_t]]
\end{eqnarray}
with real-valued time dependent functions \cite{Fuss2}
$a_t,b_t,c_t,d_t$ whose physical
meaning as drift coefficients ($a,b$) and diffusion coefficients
($c,d$) will become clear presently; they approach constant values on the 
bath correlation time scale $1/\Lambda$.

Remarkably, despite the generally non-Markovian nature of the
true open system dynamics, the evolution of the exact
$\rho_{\rm sys}(t)$ is governed by a time-local
differential equation; memory effects are encoded in the time dependent
coefficients $a_t,\ldots, d_t$. 

We need not specify the precise time dependence of
all four coefficients at this stage,
but would like to mention their behavior at early times, 
\begin{eqnarray}\label{initialbehaviour}
a_t & = &  {\cal O}(t^2),  \\ \nonumber
b_t & = &  {\cal O}(t^3),  \\ \nonumber
c_t & = & \frac{1}{2M}\langle B^2\rangle t^2 + {\cal O}(t^4) \\ \nonumber
d_t & = & \langle B^2\rangle t + {\cal O}(t^3)\,.
\end{eqnarray}
It is this early time dependence of the (diffusion)
coefficients $c_t$ and $d_t$ that will turn out
relevant for the decoherence of the largest superpositions alias
Schr\"odinger cat states, to be discussed later.

\subsection {Wigner Representation}

It is useful to switch to a phase-space representation
and to express the above master equation 
(\ref{masterosci}) for
$\rho$ as an evolution equation for the Wigner function
\be
\label{wigner}
W(q,p) = \frac{1}{2\pi\hbar}\int d\nu
\langle q-\nu/2|\rho|q + \nu/2
\rangle \e^{\i\nu p/\hbar}\,.
\ee

For the initial superposition of two Gaussian wave packets as in 
(\ref{2.1}) the Wigner function has three distinctive features as
displayed in Fig. 2: two Gaussian
wave packets in phase space arising from the diagonal terms $\rho^{11}$
and $\rho^{22}$ in (\ref{2.3}), and an oscillating pattern in between
the two Gaussians, due to the coherences $\rho^{12}$ and $\rho^{21}$.
Decoherence leads to the disappearance of those oscillations,
measured nicely by the norm $N_{12}(t)$ as we will reveal shortly.

\begin{figure}[h]\label{fig2}
\includegraphics[angle=0,scale=.80]{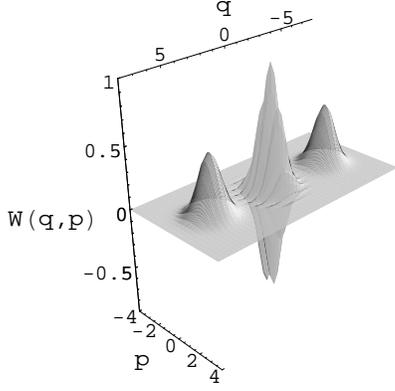}
\caption{Wigner function $W(q,p)$ of a superposition of
two Gaussian wave packets as in Fig. 1, here
with $q_1=-5=-q_2$, $p_1=p_2=0$. The
mutual distance is larger than individual spread.
The oscillating pattern in between the two Gaussians indicates 
their coherence.}
\end{figure}

In terms of the Wigner function, (\ref{masterosci})
takes the form of a
Fokker-Planck equation
\ba
\label{wignerevolution}
\dot{W}_t  & = &  \left(-\frac{\partial}{\partial q} \frac{p}{M}
+\frac{\partial}{\partial p}
\left\{(M\Omega^2 + a_t)q +b_t p\right\} \right. \nonumber  \\
 & & \left. + c_t \frac{\partial^2}{\partial p\partial q}
+ d_t \frac{\partial^2}{\partial p^2} \right)W_t\,.
\ea
We now read off the meaning of the coefficients $a,b,c,d$ in
the evolution equations: while the term involving $a_t$ is a mere
potential renormalization due to the coupling, the term involving $b_t$
describes damping. The two remaining terms represent diffusion; there is
a mixed second-order derivative with coefficient $c_t$, while the time 
dependent momentum diffusion 
involving $d_t$ reflects the stochastic force of a classical
Ohrnstein-Uhlenbeck type process. We remark that the first
diffusion term turns out to be negligible in many cases.
For a discussion of that latter 
as well as various other limits of physical relevance we refer 
the reader to \cite{masterosciref}.

We may solve equation (\ref{wignerevolution}) with the help of the
characteristic function or Fourier transform of the Wigner function,
\be\label{character}
\chi(\mu,\nu)  =  \frac{1}{2\pi\hbar} \int dq dp\;
 W(q,p) \e^{\i(\mu q -\nu p)/\hbar}\,.
\ee
Interestingly, diffusive effects on $W$ can be confined to a Gaussian
factor in $\chi$,
\begin{eqnarray}\label{gaussian}
\chi_t(\mu,\nu) & = &  \zeta_t(\mu,\nu)\exp(\int_0^t b(s)ds) \\ \nonumber
& & \times \exp\left(-\frac{1}{2\hbar^2}\left\{
\alpha_t\mu^2 + 2\beta_t\mu\nu + \gamma_t \nu^2
\right\}
\right)\,;
\end{eqnarray}
the cofactor $\zeta$ then obeys the ``Liouville'' equation
\begin{equation}\label{liouville}
\dot{\zeta}_t = \left(
-\frac{\partial}{\partial \nu}
\left(
\mu/M + b_t\nu
\right)
+\frac{\partial}{\partial \mu}
(M\Omega^2 + a_t)\nu\,\right)
{\zeta}_t\, ,
\end{equation}
provided the time dependent coefficients $\alpha,\beta,\gamma$ in the
Gaussian (\ref{gaussian}) satisfy the linear equations
\begin{eqnarray}\label{alphabetagamma}
\left( \begin{array}{c} \dot\alpha_t \\ \dot\beta_t \\
\dot\gamma_t \end{array}
\right) \!& = & \!\! 
\left( \!\begin{array}{ccc}      0       & -2/M                & 0 \\
                        M\Omega^2+a_t & -b_t               & -1/M \\
                               0       &   2(M\Omega^2+a_t) & -2b_t
\end{array}\! \right) 
\left( \!\begin{array}{c} \alpha_t \\ \beta_t \\ \gamma_t 
\end{array}\! \right) \nonumber \\ & & +
\left( \begin{array}{c} 0 \\ -c_t \\ 2d_t \end{array} \right).
\end{eqnarray}
For the transformation to ensure $\zeta_0(\mu,\nu) = \chi_0(\mu,\nu)$,
the initial condition for these coefficients is
$(\alpha_0,\beta_0,\gamma_0)=(0,0,0)$ at $t=0$. 

We may solve the ``Liouville equation'' (\ref{liouville})
through the characteristic equations
\begin{equation} \l{char}
{{\dot\mu}_t \choose {\dot\nu}_t}  =
\left({0 \atop 1/M}{-(M\Omega^2+a_t)\atop  b_t}\right)
{\mu_t \choose \nu_t }
\end{equation}
whose integral yields the linear mapping
\be\l{propagator}
{\mu_t\choose\nu_t}=M_t {\mu_0\choose\nu_0}\,;
\ee
clearly, the $2\times 2$ matrix $M_t$ originates from the $2\times2$
identity, $M_0={1\,0\choose 0\,1}$, at $t=0$.
One can further establish the identity
$ \det M_t = \exp\left(\int_0^t b(s) ds\right)\,$
which will be useful below.

Given an arbitrary initial density operator
and thus the corresponding initial characteristic function
$\chi_0(\mu,\nu)$,
we obtain the evolved $\chi_t(\mu,\nu)$
from (\ref{gaussian}), using
\begin{equation}\label{zetaprop}
\zeta_t(\mu,\nu) = \int d\mu_0 \int d\nu_0 
\chi_0(\mu_0,\nu_0) \delta(\mu-\mu_t)\delta(\nu-\nu_t),
\end{equation}
with $(\mu_t,\nu_t)$ the solutions (\ref{propagator}) of the
characteristic equations (\ref{char}).

\subsection{Coherence Norm for Distinct Wave Wackets}

The Wigner function $W$ as well as its Fourier transform $\chi$ can be
employed to express the norms $N_{ij}$ introduced in (\r{2.4}) 
and [SHB] as
\begin{eqnarray}\label{norms}
N_{ij}(t) & = & (2\pi\hbar) \int dq dp\; |W^{ij}_t(q,p)|^2 \\ \nonumber
          & = & (2\pi\hbar) \int d\mu d\nu\; |\chi^{ij}_t(\mu,\nu)|^2\,.
\end{eqnarray}
The first of these expressions nicely shows that $N_{12}(t)$ is
indeed a good indicator for the appearance of
coherences between wave packets, as it measures the weight of
the absolute square of the
oscillating pattern of the Wigner function
in between the wave packets as shown in Fig. 2.
The second expression is most convenient to actually evaluate
$N_{12}(t)$ for the oscillator model. In fact, the ansatz
(\ref{gaussian}) and the general solution (\ref{zetaprop})
yield
\begin{eqnarray}\label{n12final}
N_{12}(t) & = & (2\pi\hbar)\exp\{\int_0^t b(s)ds\} \\ \nonumber
       & &  \int\! d\mu_0 d\nu_0
\e^{-(\alpha_t\mu_t^2 +2\beta_t\mu_t\nu_t +\gamma_t\nu_t^2)/\hbar^2 }
|\chi^{12}_0(\mu_0,\nu_0)|^2
\end{eqnarray}
with $(\mu_t,\nu_t)$ the trajectories from (\ref{propagator}).

With $\rho^{12}_{\rm sys}(0) = |\varphi_1\rangle\langle\varphi_2|$ and
$|\varphi_i\ra$ representing the Gaussian wave packets (\r{2.2}),
simple Gaussian integrals give
\begin{eqnarray}\label{initialchi}
|\chi_0^{12}(\mu_0,\nu_0)|^2  & = & (2\pi\hbar)^{-2} \times \\ \nonumber 
& & \hspace{-1cm}
\exp\left\{-(\nu_0-d_Q)^2/4\sigma - \sigma
(\mu_0-d_P)^2/\hbar^2\right\},
\end{eqnarray}
where  $d_Q = |q_2-q_1|$ and $d_P = |p_2-p_1|$ denote the separations
in position and momentum of the two wave packets.
We thus see the characteristic function of the coherence to be strongly peaked
near those distances, with both widths of order $\sqrt{\hbar}$ (Recall
that we had chosen minimum uncertainty wave packets with $\sigma\sim\hbar$).

Evaluating the Gaussian integral in (\ref{n12final}) results
in the appealing form
\begin{equation}\label{finalexpression2}
N_{12}(t) = 
{\cal P}(t)
\exp\left\{-\frac{1}{\hbar^2}
(d_P,d_Q) A_t
{d_P \choose d_Q}
\right\}
\end{equation}
for the coherence norm,
revealing the quadratic dependence of the decay on the
initial separations $d_Q,d_P$. The time dependence
of this quadratic form is captured in the matrix $A_t$,
which turns out
bulky in this general case; it reads
\begin{equation}\label{amatrix}
A_t = C_t\left[
\opone +
\left({1/\sigma \atop 0}{0 \atop 4\sigma/\hbar^2}\right)C_t\right]^{-1}
\end{equation}
with the matrix
\begin{equation}\label{bmatrix}
C_t = 
 M_t^{\rm T}
\left({\alpha_t \atop \beta_t}{\beta_t \atop \gamma_t}\right)M_t,
\end{equation}
and $M_t$ the propagating matrix from (\ref{propagator}).
The functions $(\alpha_t,\beta_t,\gamma_t)$ in (\ref{bmatrix}) are
the solutions of (\ref{alphabetagamma})
with vanishing initial values
(recall that the evolution of $(\alpha_t,\beta_t,\gamma_t)$ involves
the diffusion coefficients $c_t$ and $d_t$ while only the
drift coefficients $a_t$ and $b_t$ enter the
deterministic equation for $M_t$).

A prefactor
\begin{equation}
{\cal P}(t) = \det\left(
\opone +
\left({1/\sigma \atop 0}{0 \atop 4\sigma/\hbar^2}\right)C_t\right)
^{-\frac{1}{2}}\exp\{\int_0^t b(s)ds\}
\end{equation}
appears in (\ref{finalexpression2})
which is independent of the separations and to
be identified with the similar expression (3.14-SHB) in [SHB],
resulting from the Gaussian integration.
As explained in detail in [SHB],
as long as the separation
of the two wave packets is large compared to their individual
spread, this ${\cal P}(t)$ may safely be
replaced by unity for those short times for which 
$N_{12}$ decays to essentially zero due to the relevant
exponential term in (\ref{finalexpression2}).

With the dependence of
the coherence norm on the initial
position and momentum
separations $d_Q,d_P$ now made explicit, we proceed to unveil the time
dependence of the matrix $A_t$ in the exponent 
of the coherence norm (\ref{finalexpression2}), both for the
interaction dominated early-time limit as well as for the long-time
golden-rule limit. We stress again 
that due to the large separations
$(d_Q,d_P)$ we focus on, the prefactor ${\cal P}(t)$ 
in (\ref{finalexpression2}) can and will
be replaced by unity in what follows.

\section{Limiting cases}

\subsection{Golden Rule: $\tau_{\rm sys}\ll\tau_{\rm dec}\ll\tau_{\rm diss}$}

Our exact result 
(\ref{finalexpression2}) for the coherence norm
allows us to investigate the decay of coherences on all time scales.
Matters simplify considerably in the
case of weak coupling, where decoherence
between the two wave packets may again be
evaluated analytically.

Let us first determine the 
coefficients $(\alpha_t,\beta_t,\gamma_t)$. To that end we
first recall that these coefficients vanish initially and then realize
that the inhomogeneities $c_t,d_t$ are of second order in the
interaction. We may therefore replace the $3\times 3$ propagator matrix
in the evolution equation (\r{alphabetagamma}) in zeroth order, 
i.e. by entirely neglecting
$a_t$ and $b_t$ therein. The propagator matrix is then easily
exponentiated to give
\begin{eqnarray}\label{abgperturbation}
\alpha_t & = & \int_0^t ds\; \Big(
c(s)(\sin 2\Omega (t-s))/M\Omega \\ \nonumber
& & \;\;\;\;\;\;\;\;\;\; + d(s)
(1-\cos 2\Omega (t-s))/M^2\Omega^2\Big), \\ \nonumber
\beta_t & = & -\int_0^t ds\;\Big(
     c(s) \cos 2\Omega (t-s) \\ \nonumber 
& & \;\;\;\;\;\;\;\;\;\;
+     d(s)(\sin 2\Omega (t-s))/M\Omega\Big), \\ \nonumber
\gamma_t & = & -\int_0^t ds\; \Big(
  c(s) M\Omega\sin 2\Omega (t-s) \\ \nonumber
& & \;\;\;\;\;\;\;\;\;\;
-d(s)(1+\cos 2\Omega (t-s))\Big).
\end{eqnarray}
Moreover, we may replace the time dependent coefficients
of the exact master equation (\ref{masterosci})
by their lowest-order expressions, which are
\begin{eqnarray}\label{mastercoefsperturb}
a_t & = & \int_0^t\!ds\, \la
\textstyle{\frac{\i}{\hbar}}[\tilde{B}(s),B]\ra
\cos\Omega s\;\;\;+{\cal O}(B^4)\\ \nonumber
b_t & = & \frac{1}{M\Omega}\int_0^t\!ds\, \la
\textstyle{\frac{\i}{\hbar}}[\tilde{B}(s),B]\ra
\sin\Omega s\;\;\;+{\cal O}(B^4)\\ \nonumber
c_t & = & \frac{1}{2M\Omega}\int_0^t\!ds\, \la\{\tilde{B}(s),B\}\ra
\sin\Omega s\;\;\;+{\cal O}(B^4)\\ \nonumber
d_t & = &  \int_0^t \!ds\, \textstyle{\frac{1}{2}}
\la\{\tilde{B}(s),B\}\ra\cos\Omega s\;\;\;+{\cal O}(B^4).
\end{eqnarray}
Note that perturbation theory preserves the correct
short-time expansion (\ref{initialbehaviour}).

Now that the coefficients $\alpha_t,\beta_t,\gamma_t$ within the matrix
$C_t$ in (\ref{bmatrix})
are revealed as of
second order in the interaction, it  suffices to replace $M_t$ by
its zeroth-order approximant in that exponent. Neglecting the
second-order terms $a_t$ and $b_t$ in the definition (\ref{propagator}),
we find
\begin{equation}\label{mtperturbation}
M_t = \left({\cos \Omega t \atop \sin\Omega t/M\Omega}{
-M\Omega\sin\Omega t \atop \cos\Omega t}\right)\,.
\end{equation}
With (\ref{abgperturbation}) and (\ref{mtperturbation}) we have access
to the full time dependence of the decohering quadratic form in
the exponent of the coherence norm (\ref{finalexpression2}).

Of particular interest is the long-time limit, $\Omega t \gg 1$,
which allows for many system oscillations up to the observation
time.
In that limit, furthermore assuming the bath correlation time to
be much shorter
than the system time scale $\Omega^{-1}$, we may safely replace
the time dependent coefficients $c(s)$ and $d(s)$ by their
asymptotic values $c(\infty)$, $d(\infty)$ under the
integrals in (\ref{abgperturbation}). Moreover, we see from the
latter expressions that for $\Omega t \gg 1$
all oscillating terms vanish and the only surviving contribution
arises
from the constant terms involving $d(\infty)$. The
{\it Golden-Rule limit} $\Omega t\gg 1$ thus finally yields the
coefficients
\begin{eqnarray}
\alpha_t & = & \frac{d(\infty)}{M^2\Omega^2} t,\\ \nonumber
\beta_t & = & 0, \\ \nonumber
\gamma_t & = & d(\infty) t
\end{eqnarray}
and the matrix (\ref{bmatrix})
\be\label{bmatrixGR}
C_t =  d(\infty) t
 \left({\frac{1}{M^2\Omega^2}\atop 0}{0 \atop 1}\right)\,.
\ee
With $C_t$ already of second order in the in the interaction,
we see from (\ref{amatrix}) that $A_t$ coincides
with $C_t$ in this order
of perturbation theory. The well known exponential decay
\begin{equation}\label{goldenrule}
N_{12}(t) =
\exp\left\{-\frac{d(\infty) t}{\hbar^2}
\Big(d_Q^2 + d_P^2/M^2\Omega^2
\Big) \right\}\equiv \e^{-t/\tau_{\rm dec}^{\rm GR}}
\end{equation}
characteristic of golden-rule decoherence results
(recall that we may drop the slowly varying prefactor
${\cal P}(t)$ of the general expression (\ref{finalexpression2})). 
Clearly, since
we allow the system to evolve for many cycles, any distinction of the
position as the coupling agent in the interaction Hamiltonian has
disappeared. Through the sequence of system cycles, position and
momentum interchange their role periodically such that in the long-time
limit of decoherence only the joint quantity $\left((d_Q)^2 +
(d_P)^2/(M^2\Omega^2)\right)$ appears as acceleration factor.

For the final result for the golden-rule decoherence time we
still have to determine the rate $d(\infty)$.
In the relevant order of perturbation theory and upon using the bath
correlation function (\ref{bathcorr},\r{Ohm}) we get
\begin{eqnarray}\label{asymptotic}
d(\infty) & = & \int_0^\infty\! ds\;\textstyle{\frac{1}{2}}\la
\{\tilde{B}(s),B\}\ra\cos\Omega s  \nonumber 
\\ & = & \hbar M\gamma\Omega
f_c(\Omega/\Lambda)(n_{\rm th}+\textstyle{\frac{1}{2}})\,;
\end{eqnarray}
where again
$n_{\rm th}=(\e^{\hbar\Omega/kT}-1)^{-1}$ is the thermal number of quanta
in the oscillator and $f_c(\Omega/\Lambda)\to 1$ according to
(\ref{cutoff}) . Comparing with the lowest-order drift coefficient
\begin{eqnarray}
b(\infty) & = &
\frac{1}{M\Omega}
\int_0^\infty \!ds\frac{\i}{\hbar} \la [\tilde{B}(s),B]\ra \sin\Omega s
\nonumber \\
& = & \gamma f_c(\Omega/\Lambda)\to\gamma
\end{eqnarray}
we recover the
previously announced interpretation of $\gamma$ as the classical damping
constant, $\gamma=1/\tau_{\rm diss}$, as well as the well known
golden-rule decoherence time $\tau_{\rm dec}^{\rm GR}$ already presented
in (1.2-SHB) of [SHB]. That latter expression implies the familiar
golden-rule time-scale ratio for decoherence and dissipation,
\begin{equation}
  \frac{\tau_{\rm dec}^{\rm GR}}{\tau_{\rm diss}^{\rm GR}}=
  \left(\frac{\lambda_{\rm th}}{d_{\rm eff}}\right)^2\,;
  \label{acceleration}
\end{equation}
we see decoherence accelerated over dissipation by the squared ratio of
a de Broglie wavelength $\lambda_{\rm th}=\sqrt{\hbar/M\Omega
(n_{\rm th}+1/2)}$ and an effective distance between the two
wave packets, $d_{\rm eff}=\sqrt{d_Q^2+d_P^2/M^2\Omega^2}$.

\subsection{Interaction dominance 1: 
$\tau_{\rm dec}\ll\tau_{\rm sys},\tau_{\rm diss}$}

As soon as the initial separation $(d_Q,d_P)$ between wave
packets extends beyond quantum scales, decoherence is
dominated by the interaction Hamiltonian $H_{\rm int}$ since the free 
motion generated
by $H_{\rm sys}$ eventually becomes negligibly slow by comparison.
This is the regime we discussed at length
and in a general setting in
the accompanying paper [SHB]. We must recover
those previous findings for the exactly solvable oscillator model by 
suitably simplifying
the general expression (\ref{finalexpression2}) for the coherence norm.

During these very short initial time spans,
we take into account dynamics on classical time scales
due to the ``Liouville equation'' (\ref{liouville})  to lowest order in $t$
only and replace the corresponding propagating matrix $M_t$ in 
(\ref{propagator}) by
$M_t = \left({1\atop t/M}{-M\Omega^2t \atop 1} \right) + {\cal O}(t^2)$. 
Next, we
have to determine the coefficients $\alpha_t,\beta_t$ and $\gamma_t$ 
in (\ref{alphabetagamma}).
As before, we expand all dynamical quantities connected to the
oscillator dynamics to the lowest relevant order in $t$ and find
\begin{eqnarray}\label{abgshorttime} \nonumber
\alpha_t & = & \int_0^t ds\;
 \textstyle{\langle\frac{1}{2}}\{\tilde{B}(s),B(0)\}\rangle
(2t+s)(t-s)^2/3M^2, \\  \nonumber
\beta_t & = & -\int_0^t ds\;
 \langle\textstyle{\frac{1}{2}}\{\tilde{B}(s),B(0)\}\rangle
t(t-s)/M, \\
\gamma_t & = & -\int_0^t ds\;
 \langle\{\tilde{B}(s),B(0)\}\rangle.
\end{eqnarray}
We see the importance of the details of the (real part of the)
bath correlation
function, $\langle\frac{1}{2}\{\tilde{B}(s),B(0)\}\rangle$,
describing dynamics on reservoir time scales.

The matrix (\ref{bmatrix}) 
determining the decay of coherences is
\ba\label{aud1}
C_t = \int_0^t ds\; & &
 \langle\textstyle{\frac{1}{2}}\{\tilde{B}(s),B(0)\}\rangle (t-s) \\ 
\nonumber & & \times
 \left({(t-s)(2t+s)/3M \atop t/M}\;\;\;{t/M \atop 2}\right)\, .
\ea
As in the weak coupling case, the difference between
the relevant matrix $A_t$ in (\ref{amatrix}) and $C_t$ is
negligible for the fast decoherence resulting from large
separations $(d_Q,d_P)$, and with $A_t$ replaced by
(\ref{aud1}), we get
from (\ref{finalexpression2}) 
the coherence norm as the exponential
\ba\label{gaussdecay}
N_{12}(t) &=&
\exp\,\Bigl\{
-\frac{1}{\hbar^2}
\int_0^t ds\;
 \langle\{\tilde{B}(s),B(0)\}\rangle (t-s) \\ \nonumber
& & \hspace{-.7cm}\times
\left({d_Q}^2  + d_Qd_P t/M +{d_P}^2(t-s)(2t+s)/6M^2\right) \Bigr\},
\nonumber
\ea
see \cite{BHS} and [SHB] for the case $d_P=0$.
As before, we may neglect the slowly varying prefactor ${\cal P}(t)$
of the exact expression (\ref{finalexpression2}).

Two remarks about this expression are in order. First, the oscillator
model clearly confirms the general result (6.5-SHB)
(for $d_P=0$) of the
accompanying paper [SHB], that derivation being based solely on the
assumption $\tau_{\rm dec}\ll\tau_{\rm sys},\tau_{\rm diss}$
and the Gaussian character of the bath coupling agent.
Secondly, (\ref{gaussdecay}) may be recognized to
coincide with the squared absolute norm
of the Feynman-Vernon influence functional \cite{feynman,feynman2}, with
the classical paths represented by the short-time expression
$q_t = q_1 + p_1 t/M$ (replace one by two for 
the second path $q^\prime_t$). 
Examples of decoherence following (\ref{gaussdecay})
will be shown in Sect. \ref{numerical}.

\subsection{Interaction dominance 2: 
$\tau_{\rm dec}\ll\tau_{\rm res},\tau_{\rm sys},\tau_{\rm diss}$}

In the extreme case when decoherence is even faster than
any environmental time scale it is sufficient to expand the whole matrix
$A_t$ in the exponent of (\ref{finalexpression2})
in powers of the elapsed time $t$,
mirroring the corresponding expansion of
the logarithm of the interaction propagator
in [SHB]. Crucially, we replace
the correlation function
$\frac{1}{2}\langle\{\tilde{B}(s),B(0)\}\rangle$ under the
integral by its initial value
$\langle B^2 \rangle$.
Starting from
(\ref{aud1}) of the last section we find
\begin{equation}\label{bmatrixid}
C_t = \langle B^2 \rangle \left(
{\frac{t^4}{4M^2}\atop \frac{t^3}{2M}}{\frac{t^3}{2M} \atop t^2}\right),
\end{equation}
in each entry neglecting higher order terms. Again, 
the short time approximation demands $A_t$ to be identical to
$B_t$ from (\ref{bmatrixid}) to the relevant order, and
the general result (\ref{finalexpression2}) turns into
\ba\label{shorttimeexpression}
N_{12}(t) &=&
\exp\left\{-\frac{1}{\hbar^2}\langle B^2\rangle \right. \\ \nonumber
& & \left.\left({d_Q}^2t^2 + d_Qd_P t^3/M +{d_P}^2 t^4/4M^2\right) \right\}
\nonumber
\ea
which is the universal
law (3.14-SHB) for the decay of coherences we found under very general
conditions in the accompanying paper [SHB]
and for $d_P=0$ in \cite{BHS}.
We refrain from including the prefactor ${\cal P}(t)$ here,
that factor in (3.14-SHB) being
safely replaced by unity for
the relevant short times
characteristic of the decay for asymptotically classical initial
separations $d_Q$, $d_P$. 

Any coherences present in the system will 
have vanished according to (\ref{shorttimeexpression}), before the quantum
state can be aware of any potential $V(Q)$ -- this is why
the oscillator
frequency $\Omega$ has disappeared
entirely from the final expression (\ref{shorttimeexpression})
(and also from (\ref{gaussdecay})).
Recall that system oscillations with frequency $\Omega$
were crucial for the golden-rule result.
Eq. (\ref{shorttimeexpression}) also confirms the different
scalings of the corresponding decoherence times
$\tau_{\rm dec}^Q$, $\tau_{\rm dec}^{QP}$, and
$\tau_{\rm dec}^P$ discussed in great detail in
the fourth section of [SHB].

\section{An alternative measure for decoherence}

Before turning to
actual examples,
we replace the general coherence norm
$N_{12}(t)$ by a somewhat simpler quantity with equal
power to reveal decoherence.

As we are going to restrict ourselves to
superpositions of symmetrically located wave packets with
$q_2 = -q_1$, and $p_2=-p_1$, an alternative
indicator of the decay of coherences is
the value of the Wigner function of $\rho_{\rm sys}$
at the origin, 
\be
W_t(0,0)  =  \frac{1}{\pi\hbar} \int dq \la q|\rho| -q\ra 
=  \frac{1}{\pi\hbar} \int dp \la p|\rho| -p\ra.
\ee
Apparently, $W_t(0,0)$ is a measure for the weight off the diagonal,
both in the position and momentum representations; it is linear in the
density operator and hence easier to access numerically, given the 
master equation; 
it is intimately related to the previously employed $N_{12}(t)$ as we 
will briefly show.
By definition, we have 
$W_t(0,0) = \frac{1}{2\pi\hbar}\int d\mu d\nu \chi_t(\mu,\nu)
= \frac{1}{2\pi\hbar}\int d\mu d\nu \sum_{ij}c_ic_j^*\chi_t^{ij}(\mu,\nu)$.
The diagonal terms $\chi_t^{11}$ and $\chi_t^{22}$ do not noticeably contribute
to $W_t(0,0)$ since for superpositions of far-apart wave packets their 
Wigner correspondant
is located near $(q_1,p_1)$, and $(-q_1,-p_1)$, respectively,
i.e. far away from the phase space origin, as is also
apparent in Fig. 2.
Furthermore, for the symmetric case considered here, we find
real Gaussians for the characteristic functions
such that $\chi_t^{12}=\chi_t^{21} = |\chi_t^{21}|$ and therefore
\begin{equation}\label{w00}
W_t(0,0) \propto
\int d\mu d\nu |\chi_t^{12}(\mu,\nu)|\,.
\end{equation}
Comparing the latter with expression (\ref{norms}) for the 
norms $N_{ij}$
we see that for the symmetric
superposition of wave packets considered here, the linear 
quantity $W_t(0,0)$
has equal power to reveal the fate of coherence as the
coherence norm $N_{12}(t)$.
Going through the very same steps as in the previous 
section we find
\begin{eqnarray}\label{finalw00}
n_{12}(t) & \equiv &  W_t(0,0)/W_0(0,0) \\ \nonumber & = & 
{\cal P}(t)
\exp\left\{-\frac{1}{2\hbar^2}
(d_P,d_Q) A_t
{d_P \choose d_Q}
\right\}\, ,
\end{eqnarray}
with the same prefactor
${\cal P}(t)$ and matrix $A_t$ as in
the expression (\ref{finalexpression2})
for $N_{12}(t)$. Neglecting the irrelevant prefactor 
${\cal P}(t)$ reveals that the simpler quantity
$n_{12}(t)$ is essentially just the square root of the coherence
norm $N_{12}(t)$ and our previous discussion of that
quantity in Sect. \ref{Secthomodel} equally
applies to $n_{12}(t)$.

\section{Numerical results for the Oscillator model}\label{numerical}

We illustrate
numerically the decoherence of a superposition of two
symmetrically located coherent states, 
$|\varphi_1\ra = |\alpha\rangle$ and
$\varphi_2 = |-\alpha\rangle$, which are labeled
by a complex number $\alpha$,
$|\alpha\rangle = \e^{-\frac{1}{2}|\alpha|^2} \e^{\alpha a^+} |0\rangle$,
see \cite{Glauber}.
The dimensionless creation and annihilation operators, for instance
$a  = \left(\sqrt{M\Omega/2\hbar}\;Q + \i P/\sqrt{2M\Omega\hbar}\right)$,
contain the quantum reference scales of length and 
momentum characteristic of the harmonic oscillator.
We may employ those scales to introduce dimensionless distances 
between the two coherent states
\begin{eqnarray}\label{dimless}
{\tilde d}_Q & = & d_Q/\sqrt{\hbar/M\Omega} = \sqrt{2}(\alpha + \alpha^*)
\\ \nonumber
{\tilde d}_P & = & d_P/\sqrt{M\hbar\Omega} =  \sqrt{2}(\alpha - \alpha^*)/\i\,,
\end{eqnarray}
which will be the relevant quantities in the following. In these units, 
large separations with respect to the spread of coherent states,
as required for our general formula (\ref{finalexpression2}),
simply means large compared to unity. Truly macroscopic separations
could lead to dimensionless separations of the order, say, 
${\tilde d}_Q \simeq 10^{17}$, a scale way beyond our examples below. 
We shall let
${\tilde d}_Q$ or ${\tilde d}_P$ range from $16$ to $4000$ and in one 
extreme case
to $800000$. It will become clear that the transition from slow golden-rule 
decoherence to 
rapid interaction dominated decoherence is fully captured in the 
mesoscopic range considered.

We assume a thermal environment with all parameters
of the total Hamiltonian fixed once and for all.
It is the initial state of the
system oscillator only that we alter between simulations. 
The transition we are after is traced out by varying the dimensionless 
initial phase space
distances $({\tilde d}_Q, {\tilde d}_P)$
between the superposed wave packets.

All simulations are performed with the small damping rate
$\gamma = 10^{-5}\Omega$, a rather large 
temperature according to $k_BT = 20\hbar\Omega$, and an environmental 
cutoff frequency
$\Lambda = 100\Omega$. The latter is the largest frequency involved
in either system or environment. We expect interaction dominated decoherence 
according to (\ref{gaussdecay}) 
to set in as soon as the decoherence time
scale becomes shorter than the shortest system time scale, here $\Omega^{-1}$.
The extreme case of quadratic decay 
according to (\ref{shorttimeexpression})
(or qubic or quartic, depending on
the nature of the superposition - see also [SHB])
occurs as soon as $\tau_{\rm dec}$
is even shorter than reservoir time scales, here
determined by $\Lambda^{-1}$.
The numerical solution of the exact master equation
is based on the corresponding exact stochastic 
Schr\"odinger equation for this model, as
established recently \cite{Stochastic}. This approach
has the nice feature of preserving Gaussian pure states within each run of the
Schr\"odinger equation for a given realization of the driving noise
and, moreover, allows easy access to the otherwise
involved expressions for the time dependent
coefficients entering the evolution equations.

\subsection{Golden rule}

We can estimate for which dimensionless effective
separation 
${\tilde d}_{\rm eff} = \sqrt{{\tilde d}_Q^2 + 
{\tilde d}_P^2}$
the golden rule result should apply
for our choice of parameters.
>From (\ref{goldenrule}) and
(\ref{asymptotic}) we find
$\tau_{\rm dec}^{\rm GR} = 
\left(\gamma(n_{\rm th}+1/2){\tilde d}_{\rm eff}^2\right)^{-1}$.
Recall that the golden rule assumes a long-time limit
$\Omega t \gg 1$. We can thus trust the golden rule 
decay law only as long as 
$\Omega \tau_{\rm dec}^{\rm GR} \gg 1$ which puts an upper bound
on ${\tilde d}_{\rm eff}$. Together with the lower bound arising from the 
requirement of a large separation we have 
\begin{equation}
1 \ll {\tilde d}_{\rm eff} \ll \sqrt{\Omega/\gamma(n_{\rm th}+1/2)}\,.
\end{equation}
For the choice of parameters above, we find
$1 \ll {\tilde d}_{\rm eff} \ll 70$, a fairly narrow band
for the applicability of the golden rule.

\vspace*{0.5cm}
\begin{figure}[h]\label{fig3}
\includegraphics[angle=0,scale=.30]{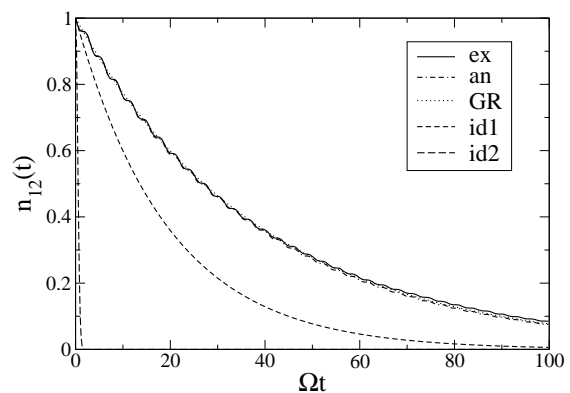}
\caption{
Golden-rule decay of an initial superposition of
two coherent states. 
Distances are
${\tilde d}_Q=16$ and ${\tilde d}_P=0$. Environmental
parameters are $\gamma=10^{-5}\Omega$, $\Lambda = 100\Omega$,
and $kT=20\hbar\Omega$.
Full line (``ex''): exact numerical calculation, indistinguishable from 
the analytical expression
(\ref{finalexpression2})
(dash-dotted, ``an'').
Dotted line (``GR''): golden-rule
result (\ref{goldenrule}) which fits smoothly through the
slightly oscillatory exact result.
Interaction-dominance results (dashed lines, ``id1'', ``id2'') 
are irrelevant here;
the influence of the system Hamiltonian is crucial.}
\end{figure}

In Fig. 3 we show the decay of an initial superposition of
two coherent states with separations
${\tilde d}_{\rm eff} = {\tilde d}_Q = 16$ and thus
${\tilde d}_P=0$. The full line (``ex'') shows the exact numerical
result. Apart from some tiny oscillations around the exponential decay 
the golden-rule expectation (dotted and labeled ``GR'') 
according to (\ref{goldenrule}) 
is nicely confirmed. Note how the tiny oscillations loose
final relevance as soon as $\Omega t\gg 1$. These oscillations have their 
origin in the periodic change of the system state between a 
superposition with respect to $Q$ to one with respect to $P$.
Decoherence is more efficient for $Q$-superpositions,
and less effective in those time spans when the system state assumes
a superposition with respect to $P$. During
those periods decoherence is slowed down slightly.
During one oscillator period there are two swaps between a $Q$- and a 
$P$-superposition, which fact explains the frequency $2\Omega$ of the 
oscillations in question. The dash-dotted line (``an'') depicts
the analytical expression (\ref{finalexpression2}) and is almost 
indistinguishable
from the exact result. The two dashed curves (``id1'', ``id2'')
correspond to predictions from the two
interaction dominated regimes. They are clearly irrelevant for
the still ``slow'' decoherence of the underlying ``microscopic'' superposition.
The system Hamiltonian has a strong influence on the decay in that regime.

\subsection{Crossover regime: $\tau_{\rm dec}\approx\tau_{\rm sys}$}

The golden rule ceases to be relevant as soon as the initial separation
between the superposed wave packets increases to such an
extent that the decoherence
time scale $\tau_{\rm dec}$ becomes comparable to
$\tau_{\rm sys}$.
Then the decay must still be system specific, that is 
depend on the details of the dynamics induced by
the system Hamiltonian and the time scales emerging from
the bath correlation function. It is the special virtue of the 
oscillator model to allow rigorous treatment of this crossover case. 

\vspace*{0.5cm}
\begin{figure}[h]\label{fig4}
\includegraphics[angle=0,scale=.30]{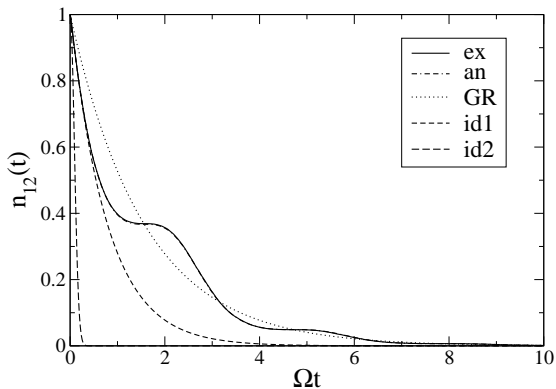}
\caption{
Crossover regime, ${\tilde d}_Q=80$ and ${\tilde d}_P=0$, otherwise
same parameters as in Fig.~3. Golden rule no longer valid,
universal interaction dominated regime not yet reached.
Full line (``ex'')for exact numerical calculation;
indistinguishable dash-dotted line (``an'') for (\ref{finalexpression2}). 
Only for small times, $\Omega t\ll 1$, decay according to 
(\ref{gaussdecay}) (``id1'') becomes visible.
Oscillations due to system Hamiltonian, see main text.}
\end{figure}

Fig. 4 shows the decay of an initial superposition of
coherent states with an intermediate position separation
${\tilde d}_Q = 80$ and equal momenta,
${\tilde d}_P=0$. The (indistinguishable) full and dash-dotted lines
show the exact master equation
result and the evaluation of the exact analytical
expression (\ref{finalexpression2}).
We see that decoherence due to coupling to 
the position is very effective initially; however, as the
system dynamics moves the $Q$-superposition to
one with respect to $P$ after a quarter of the oscillator
period $\Omega t \approx \pi/2$, decoherence slows down. 
Turning the state further in phase space, 
decoherence becomes efficient again after another quarter of the
period, and periodically thereafter. The periodic alteration between
slowdown and acceleration is now a prominent feature rather than a 
small fluctuation. For times short compared to the
system time scale, $\Omega t\ll 1$, the influence of the
system Hamiltonian is negligible and thus the
general result (\ref{gaussdecay}) of interaction dominance 
(dashed line, ``id1'') initially
applies as can be seen from the good agreement
for short times. 

\subsection{Interaction dominated decoherence 1}

As soon as the separations
${\tilde d}_Q$ or ${\tilde d}_P$ 
between the superposed coherent states are sufficiently
large, the system Hamiltonian becomes irrelevant and
interaction dominates, see Fig. 5. For a separation
of ${\tilde d}_Q = 200$ and equal momenta
${\tilde d}_P=0$, decoherence is faster than any
system time scale, yet still longer than the environmental
correlation time $\Lambda^{-1}$. This is the regime
that can fully be described by the general 
decay law (6.5-[SHB]) of the accompanying paper [SHB], here
(\ref{gaussdecay}).
As mentioned in [SHB], unless the spectral density differs
from zero at zero frequency -- which is not the case here,
(\ref{gaussdecay}) describes non-exponential decay.
We further remark that the golden rule (dotted line, ``GR'')
now predicts far too slow decoherence.

\vspace*{0.5cm}
\begin{figure}[h]\label{fig5}
\includegraphics[angle=0,scale=.30]{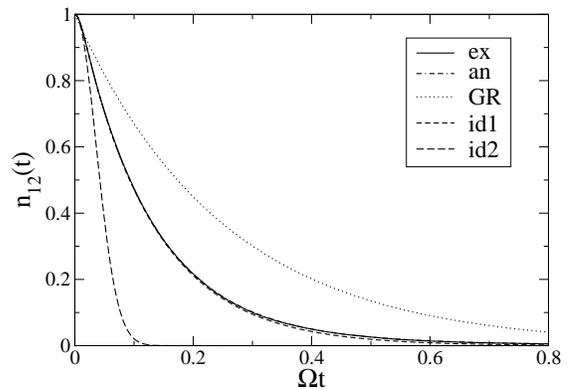}
\caption{
Interaction dominance, ${\tilde d}_Q=200$ and ${\tilde d}_P=0$, otherwise
same as in Fig.~3; note change of time scale.  
Full line (``ex'') for exact numerical calculation,
indistinguishable dash-dotted line (``an'') for (\ref{finalexpression2}). 
Golden
rule (dotted line, ``GR'') now meaningless; interaction dominance
(dashed line, ``id1'')  throughout whole
relevant time span. Extreme short-time result (\ref{shorttimeexpression}) 
(dashed line, ``id2'') enters validity for very early times.}
\end{figure}

\subsection{Interaction dominated decoherence 2}

We expect universal decay according to our simple analytical
expression (\ref{shorttimeexpression})
(corresponding to result (3.14-[SHB]) in [SHB])
as soon as the decoherence time is
the shortest time scale involved, here even shorter than the inverse
environmental cutoff frequency $\Lambda^{-1}$. We thus find lower
bounds for the size of superpositions that indicate the
beginning of this 
universal interaction dominated decoherence regime.

We use (\ref{bquadrat}) to find an expression for
$\la B^2\ra$ and conclude from the discussion of
interaction dominated decoherence times in [SHB] that
for a superposition in position space, 
$\Lambda \tau_{\rm dec}^{Q} \simeq
\sqrt{\hbar\Lambda\Omega/(kT\gamma)}/{\tilde d}_Q$.
The latter has to be small compared to one in order for this
universal regime to be important. With the choice of parameters
as before, we find the condition ${\tilde d}_Q \gg 700$ for the
universal Gaussian decay law to be relevant.

\vspace*{0.5cm}
\begin{figure}[h]\label{fig6}
\includegraphics[angle=0,scale=.30]{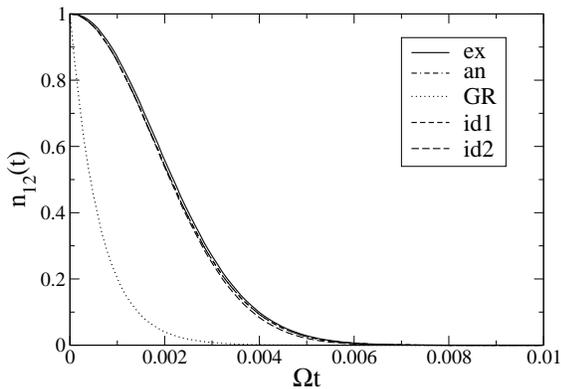}
\caption{
Interaction dominated universal decoherence,
${\tilde d}_Q=4000$ and ${\tilde d}_P=0$, otherwise as 
in Fig.~3; note change of time scale.
Four results in agreement:
full line (``ex'') for exact numerical calculation,
dash-dotted line (``an'') for analytical
formula (\ref{finalexpression2}), two dashed
lines (``id1'' and ``id2'') for interaction dominance, the 
latter the Gaussian
$\exp\{-(t/\tau_{\rm dec}^Q)^2\}$ from (\ref{shorttimeexpression}).
Golden rule (dotted line, ``GR'') grossly underestimates life 
time of coherence.}
\end{figure}

In Fig. 6 we show the decay of an initial superposition of
coherent states with a yet larger separation
${\tilde d}_Q = 4000$ and 
${\tilde d}_P=0$.
Note the difference in time scales along the $t$-axis between the
various figures.
While the full line (``ex'') represents the exact numerical calculation,
the dash-dotted line (``an'') is our exact
formula (\ref{finalexpression2}). In this case,
interaction dominance holds
(dashed lines, ``id1'' and ``id2''), the
extreme short-time result 
(\ref{shorttimeexpression}) suffices to describe the decay
over the full relevant time interval.
The latter is a Gaussian
$\exp\{-(t/\tau_{\rm dec}^Q)^2\}$ decay law.
Clearly, the golden rule is not applicable here and wrongly
predicts far too fast decoherence. This is due to the
quadratic golden rule scaling of the decoherence times with distance,
as compared to only a linear dependence of the true interaction
dominated decoherence time [SHB].

A superposition in momentum space with equal
positions ${\tilde d}_Q = 0$ is more stable under the
position dependent
coupling $H_{\rm int}=QB$ between system and environment. To reveal
interaction dominance and
our universal decay law, a momentum superposition will have to
be much larger (in our dimensionless units) than a
position space superposition.
As in the previous case,
we get an estimate for the border of applicability of
our universal law (see the discussion in the accompanying paper [SHB])
$\Lambda \tau_{\rm dec}^{P} \simeq
\left(\hbar\Lambda^3/(kT\Omega\gamma)\right)^{\frac{1}{4}}/
\sqrt{{\tilde d}_P}$.
This quantity has to be small compared to one for
universal ``quartic'' decay.
For our choice of parameters, ${\tilde d}_P \gg 70000$ is required.

\vspace*{0.5cm}

\begin{figure}[h]\label{fig7}
\includegraphics[angle=0,scale=.30]{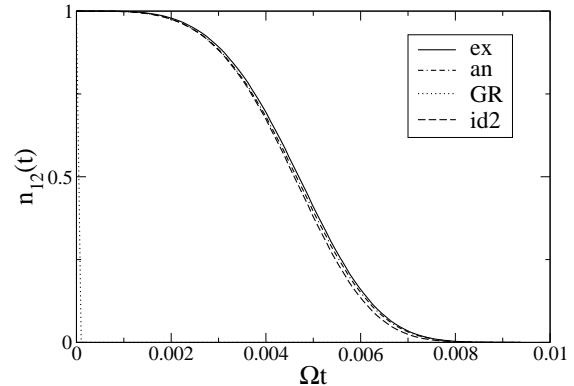}
\caption{
Interaction dominated universal decay of initial 
superposition of two coherent states with opposite
momenta and equal positions.
Distances are
${\tilde d}_Q=0$ and ${\tilde d}_P=800000$, other 
parameters as in Fig. 3; note change of time scale.
Four results in agreement:
full line (``ex'') for exact numerical calculation,
dash-dotted line (``an'') for analytical
formula (\ref{finalexpression2}), and a dashed
line (``id2'') of interaction dominance, the 
latter now the quartic exponential
$\exp\{-(t/\tau_{\rm dec}^P)^4\}$ law (\ref{shorttimeexpression}).
Golden rule (dotted line ``GR'') underestimates lifetime of coherence
even more drastically as in Fig.~6.}
\end{figure}

We check on that case in Fig. 7 which depicts the decay of an initial 
superposition of two coherent states with momentum separation
${\tilde d}_P = 800000$ and equal positions
${\tilde d}_Q=0$.
As before, the full line (``ex'') is the exact numerical
result compared with
(\ref{finalexpression2}) (dash-dotted line)
and the interaction dominated results
(dashed line, ``id2'').
Again, the golden rule by far underestimates the decoherence time,
due to the wrong scaling: while
$\tau_{\rm dec}^{GR}$ scales with 
${\tilde d}_P^2$, the true
$\tau_{\rm dec}^{P}$, as it may be determined from our
approach to interaction dominated decoherence [SHB], scales
with the square root $\sqrt{{\tilde d}_P}$ only.
However, Fig. 7 does confirm the quartic decay
$n_{12}(t)=\exp\{-(t/\tau_{\rm dec}^P)^4\}$ we predict according to
(\ref{shorttimeexpression}). 

\section{Conclusions}

The exactly soluble model of a damped harmonic
oscillator coupled to bath of oscillators illustrates
our prediction of various decoherence scenarios.
For microscopic distances between the superposed
wave packets, decoherence is so slow as to witness many free-system cycles
and follows the standard exponential golden-rule decay. As the
separation increases, decoherence and system time scales eventually become
comparable and a more complicated, system-specific decay of
coherences results. 
Our findings indicate that golden-rule predictions
may fail for even only moderately sized superpositions.
For further increased separations we enter
interaction dominated decoherence as presented in \cite{BHS}
and in much more detail in the accompanying paper [SHB].
Now $\tau_{\rm dec}\ll\tau_{\rm sys}$ and decoherence
proceeds according to a system-independent decay law
(\ref{gaussdecay}), incorporating the bath correlation
function.
In the extreme case when decoherence even outruns
intra-bath processes, the decay becomes 
a simple Gaussian for an initial separation
with respect to position, and an initially slower fourth-order exponential
if momentum is the differing property; position is distinguished over 
momentum since it is taken as the system coupling agent in the interaction
with the bath.

The special virtue of the oscillator model studied here is to allow
inspection of the crossover between the
long-time golden rule regime valid for microscopic
superpositions, and the interaction dominated regime
relevant for more and more macroscopic superpositions.

As experimental efforts towards resolving decoherence dynamics
continue \cite{experiments}, it will be fascinating to see
whether larger-scale superpositions can be
realized and whether decoherence dynamics can be pushed towards
the interaction dominated regime illustrated here
and in [SHB].

\section*{Acknowledgments}

We have enjoyed discussions with Wojciech Zurek, Daniel Braun,
Dieter Forster, 
and Hans Mooij as well as
the hospitality of the Institute for Theoretical Physics at the
University of Santa Barbara during the workshop
'Quantum Information: Entanglement, Decoherence, Chaos', 
where this project was completed.
Support by the Sonderforschungsbereich
``Unordnung und Gro{\ss}e Fluktuationen'' (Essen) and
``Korrelierte Dynamik hochangeregter atomarer und molekularer Systeme''
(Freiburg)
of the Deutsche Forschungsgemeinschaft is gratefully acknowledged.


\begin{references}

\bibitem[SHB]{SHBPRA} W. T. Strunz, F. Haake, and  D. Braun,
                  Phys. Rev. A. (accompanying paper).

\bibitem{generalosci}
J. R. Senitzky, Phys. Rev. {\bf 119}, 670 (1960);
F. Schwabl, W. Thirring, Ergeb. exakt.  Naturwiss. {\bf 36}, 219 (1964); 
G. W. Ford, M. Kac, and P. Mazur, J. Math. Phys. {\bf 6}, 504 (1965);
P. Ullersma, Physica {\bf 32}, 27 (1966); 
A.O. Caldeira, A.J. Leggett, Phys. Rev. {\bf A31}, 1059 (1985);
F. Haake, M.  {\.Z}ukowski, ibid. {\bf A47}, 2506 (1993).

\bibitem{masterosciref}
F. Haake, R. Reibold, Phys. Rev. {\bf A32}, 2462 (1985);
B. L. Hu, J. P. Paz, and Y. Zhang, Phys. Rev. {\bf D45}, 2843 (1992).

\bibitem{Zeh} D. Giulini, E. Joos, C. Kiefer, J. Kupsch, I.O. Stamatescu,
H.D. Zeh, {\it Decoherence and the appearance of a classical world in
quantum theory} (Springer, Berlin 1996).

\bibitem{Zurek} W.H. Zurek, Phys. Today {\bf 44}, 36-44 (1991); W.H.
Zurek, S. Habib, J.P. Paz, Phys. Rev. Lett. {\bf 70}, 1187 (1993); J.P.
Paz, W.H. Zurek, Phys. Rev. Lett. {\bf 82}, 5181 (1999)

\bibitem{Habib} J. P. Paz, S. Habib, W. H. Zurek, Phys. Rev. {\bf D47},
  488 (1993).

\bibitem{Glauber} R.J. Glauber, Phys. Rev. {\bf 131} 2766 (1963);
C.G. Sudarshan, Phys. Rev. Lett. {\bf 10}, 277 (1963).

\bibitem{feynman} 
R.P. Feynman, F.L. Vernon, Ann. Phys. (N.Y.) {\bf 24}, 118 (1963).

\bibitem{unruhetc}
W. G. Unruh and W. H. Zurek, Phys. Rev. {\bf D40}, 1071 (1989);
G. W. Ford and R. F. O'Connell, Phys. Rev. {\bf D64}, 105020 (2001).
 
\bibitem{Weiss} U. Weiss, {\it Quantum dissipative systems},
(World Scientfic, Singapore), 2nd edition, (2000).

\bibitem{Fuss2} To save a little
space we shall in the sequel mostly write the time argument as an index
and even sometimes drop the latter.

\bibitem{BHS} D. Braun, F. Haake, and W. T. Strunz, 
                  Phys. Rev. Lett. {\bf 86}, 2913 (2001).

\bibitem{feynman2}
We thank D. Cohen for drawing our attention to this connection.

\bibitem{Stochastic} 
L. Diosi, N. Gisin, and W. T. Strunz, Phys. Rev {\bf A58}, 1699 (1998);
W. T. Strunz, Chem. Phys. {\bf 268}  237 (2001).

\bibitem{experiments} M. Brune, Phys. Rev. Lett. {\bf 77}, 4887 (1996);
C.J. Myatt, B.E. King, Q.A. Turchette, C.A. Sackett, D.
Kielpinski, W.M. Itano, C. Monroe, D.J. Wineland, Nature {\bf 403}, 269
(2000).


\end{references}
\end{document}